\documentclass{webofc}
\usepackage[varg]{txfonts}
\usepackage{epsfig}
\usepackage{amssymb}
\usepackage[figuresright]{rotating}
\begin{document}
\title{Fluctuating fluid dynamics for the QGP in the LHC and BES era}
\author{\firstname{Marcus} \lastname{Bluhm}\inst{1,\,2}\fnsep\thanks{\email{marcus.bluhm@uwr.edu.pl}}
        \and
        \firstname{Marlene} \lastname{Nahrgang}\inst{\,2}
        \and
        \firstname{Thomas} \lastname{Sch\"afer}\inst{\,3}
        \and
        \firstname{Steffen~A.} \lastname{Bass}\inst{\,4}
}
\institute{Institute of Theoretical Physics, University of Wroc\l{}aw, 50204 Wroc\l{}aw, Poland
\and
           SUBATECH UMR 6457 (IMT Atlantique, Universit\'e de Nantes, IN2P3/CNRS), 4 rue Alfred Kastler, 44307 Nantes, France
\and
           Physics Department, North Carolina State University, Raleigh, NC 27695, USA
\and
           Department of Physics, Duke University, Durham, NC 27708-0305, USA
}
\abstract{In an era of high-precision determinations of QGP properties a full incorporation of fluid dynamical fluctuations into 
our models has become crucial, in particular, when describing the dynamics of small systems or near the conjectured QCD critical 
point. In this talk we discuss some effects of the propagation of these fluctuations. For LHC physics we focus on fluctuations 
in the energy-momentum tensor, while the impact of fluctuations in the diffusive net-baryon density is studied to improve our 
knowledge on the formation of critical fluctuations being searched in current and future BES programs.}
\maketitle
%
\section{Introduction}
\label{intro}

The bulk expansion of the quark-gluon plasma (QGP) formed in relativistic heavy-ion 
collisions~\cite{Shuryak:1980tp,Heinz:2005zg} can successfully be described by means of conventional dissipative fluid 
dynamics~\cite{Romatschke:2009im}. In conventional fluid dynamics thermal averages of energy, momentum and conserved charges, 
such as net-baryon number, are propagated. Including dissipative effects into the fluid dynamical framework demands, however, to also 
incorporate fluctuations. This is because the role of the dissipation, in accordance with the fluctuation-dissipation theorem, is to 
counteract intrinsic fluid dynamical fluctuations which drive systems constantly out of equilibrium. These depend on the number of 
particles or the volume and, thus, become essential for small systems or systems undergoing a phase transition where long-range 
correlations develop. 

In the vicinity of the conjectured QCD critical point fluctuation signals, as searched in the beam energy scan (BES) programs, are 
expected to be significantly enhanced~\cite{Stephanov:1998dy,Stephanov:1999zu,Stephanov:2008qz,Asakawa:2009aj}. The diffusive 
net-baryon density $n_B$ becomes the slowest critical mode~\cite{Son:2004iv,Fujii:2004jt}, and fluctuations in the net-baryon number 
and related observables promise therefore indicative signatures. Nonequilibrium effects caused by the fast expansion of the matter 
can, however, quantitatively affect the fluctuation signals 
\cite{Berdnikov:1999ph,Nahrgang:2011mg,Nahrgang:2011mv,Nahrgang:2011vn,Herold:2013bi,Kitazawa:2013bta,Mukherjee:2015swa,Herold:2016uvv}. 
This talk reports, in continuation of~\cite{Nahrgang:2017oqp}, on our recent progress in developing dynamical models capable of fully 
propagating intrinsic fluctuations in fluid dynamics as relevant both for vanishing $n_B$ (Sec.~\ref{sec-1}) and for the dynamics of 
critical fluctuations (Sec.~\ref{sec-2}). 

\section{The role of fluid dynamical fluctuations}
\label{sec-1}

Fluid dynamical fluctuations are embedded into the full evolution equations of a relativistic fluid via 
\begin{equation}
 \partial_\mu T^{\mu\nu}=\partial_\mu\left(T^{\mu\nu}_{\rm eq} + \Delta T^{\mu\nu}_{\rm visc} + \Xi^{\mu\nu}\right) = 0 
 \label{eq:fluideq}
\end{equation}
by adding the noise field tensor $\Xi^{\mu\nu}$ to the conventional dissipative energy-momentum tensor $T^{\mu\nu}$. The correlators 
of the noise components are obtained in linear response theory and read for a conformal fluid in $3+1$ dimensions in Gaussian white 
noise approximation 
\begin{equation}
 \langle\Xi^{\mu\nu}(x)\,\Xi^{\alpha\beta}(x')\rangle = 2T\eta\left(\Delta^{\mu\alpha}\Delta^{\nu\beta}+\Delta^{\mu\beta}
 \Delta^{\nu\alpha}-\frac23\Delta^{\mu\nu}\Delta^{\alpha\beta}\right)\delta^{(4)}(x-x') 
 \label{eq:covariance}
\end{equation}
while their mean values vanish. For relativistic fluids the dissipative corrections $\Delta T^{\mu\nu}_{\rm visc}$ are dynamical 
quantities whose evolution is subject to Israel-Stewart type relaxation equations. It makes sense to evolve the noise fields in a 
similar fashion via $u^{\gamma}\partial_\gamma\Xi^{\langle\mu\nu\rangle}=-\,(\Xi^{\,\mu\nu}-\xi_{\rm Gauss}^{\,\mu\nu})/\tau_\xi$, 
where the components $\xi_{\rm Gauss}^{\,\mu\nu}$ are correlated as in Eq.~(\ref{eq:covariance}). As a consequence, the noise is 
correlated in time within the relaxation time $\tau_\xi$. 

We have incorporated fluctuations in $T^{\mu\nu}$ into the $3+1$d relativistic viscous fluid dynamics code vHLLE~\cite{Karpenko:2013wva}. 
For exemplary numerical studies we choose the noise relaxation time to be the same as that of the shear stresses and consider a fluid in a 
static box with periodic boundary conditions. With decreasing lattice spacing $\Delta x<1$~fm the handling of large gradients caused by 
the increasing variance of the noise becomes challenging for the present algorithm such that we opt for smoothing the noise fields over 
lengths of $1$~fm. 

\begin{figure}
 \centering
 \sidecaption
 \includegraphics[width=0.55\textwidth]{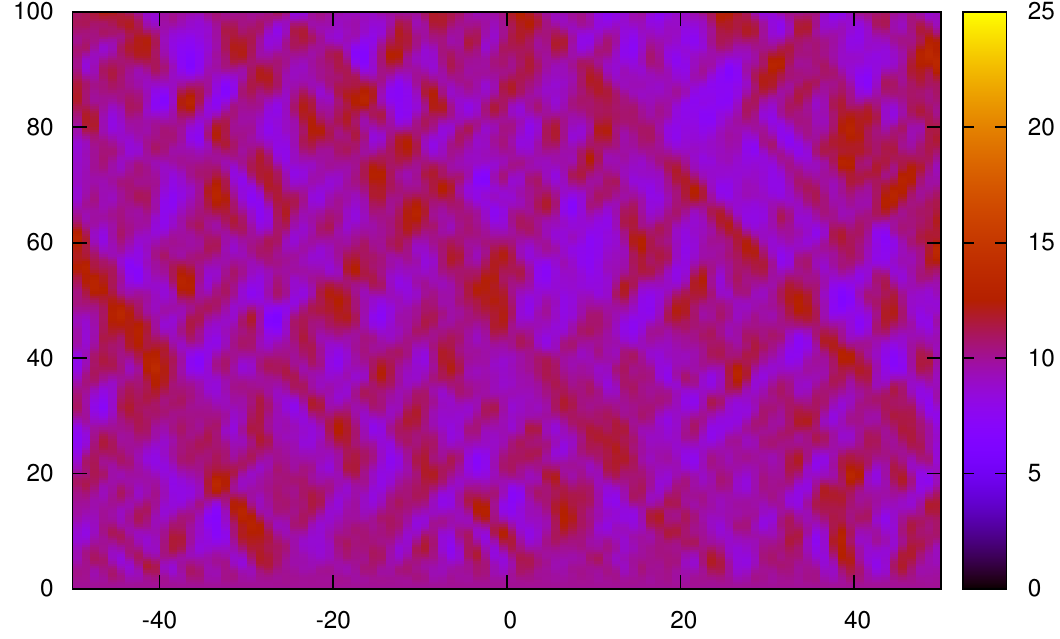}
 \caption{Spatio-temporal evolution of the energy density in units of GeV/fm$^3$ shown in one spatial direction within the first 
 $100$~fm/c for a fluid confined in a box of length $100$~fm. The fluid has a specific shear viscosity $\eta/s=0.2$ and initially a 
 homogeneous energy density $e_0=10$~GeV/fm$^3$. This result is obtained for a resolution of $1/\Delta x=1/$fm.}
 \label{fig:edenscolormap}
\end{figure}
Figure~\ref{fig:edenscolormap} shows the evolution of the energy density, which we initialize at a homogeneous value $e_0$, along one 
spatial direction. Fluctuations induce local deviations from the average energy and momentum densities which are propagated by the fluid 
dynamical equations. As time progresses, a non-zero local variance of the energy density builds up and saturates. Moreover, the average energy 
density in the box is reduced by about 3\% at late times. This is because the nonlinearities in fluid dynamics modify thermal averages in a 
nontrivial way: the interactions of the fluctuations lead, for example, to cutoff-dependent corrections in the equation of state and the 
transport coefficients~\cite{Kovtun:2011np,Chafin:2012eq}. A diligent numerical implementation of fluctuating fluid dynamics needs to 
correct such effects. This will be a crucial aspect of future work on small systems as created in p-p(A) collisions~\cite{Yan:2015lfa}.

\section{Stochastic diffusion of critical fluctuations}
\label{sec-2}

Similarly, fluctuations can be included into the conservation equation of the net-baryon number 
$\partial_\mu N^\mu=\partial_\mu\left(N^\mu_{\rm eq} + \Delta N^\mu_{\rm visc} + I^\mu\right) = 0$ via a noise vector $I^\mu$. 
Considering, for simplicity, a nonrelativistic fluid with space-time independent fluid velocity at a homogeneous temperature $T$, the 
conservation equation reduces to a stochastic diffusion of the net-baryon density, 
\begin{equation}
 \partial_t n_B(t,x) = \Gamma\,\nabla^2{\cal F}'[n_B] + \nabla J(t,x) \,,
 \label{eq:diffeq}
\end{equation}
which follows the minimization of the free energy functional 
\begin{equation}
 {\cal F}[n_B] = T\int{\rm d}^3 x \left(\frac{m^2}{2n_c^2}\left(\Delta n_B\right)^2 + \frac{K}{2n_c^2}\left(\nabla n_B\right)^2\right) \,.
 \label{eq:freeEfunctional}
\end{equation}
Here, $\Gamma$ is the mobility coefficient, $\Delta n_B=n_B-n_c$ with critical density $n_c$, and $J(t,x) = \sqrt{2T\Gamma}\zeta(t,x)$ is 
a stochastic current with Gaussian white noise $\zeta(t,x)$, which itself has unit variance. We include criticality via the mass term, 
$m^2=1/(\xi_0\xi^2)$, which is a function of the equilibrium correlation length $\xi$ whose temperature-dependence can be obtained from the 
$3$d Ising model equation of state~\cite{Bluhm:2016byc,Nahrgang:2017hkh}. The kinetic term in Eq.~(\ref{eq:freeEfunctional}) depends on the 
surface tension $K$. 

The free energy functional contains only Gaussian terms such that Eq.~(\ref{eq:diffeq}) is linear in $n_B$. This allows us to study the 
stochastic equation numerically with a simple implicit scheme. Moreover, we confine the diffusion to a one-dimensional system of length 
$L$ with periodic boundaries which implies that cutoff-effects are absent and the results approach continuum expectations with decreasing 
lattice spacing. In the numerics we initialize $n_B$ homogeneously at $n_c=1/(3$fm$^3)$, choose $\Gamma=1/(3T$fm$^2)$ and $\xi_0=0.48$~fm, 
fix the pseudo-critical temperature $T_c=0.15$~GeV at which $\xi(T_c)\simeq 6.4\,\xi_0$, and analyze equilibrium observables only after a 
sufficient equilibration time. 

\begin{figure}[t]
 \centering
 \includegraphics[width=0.48\textwidth]{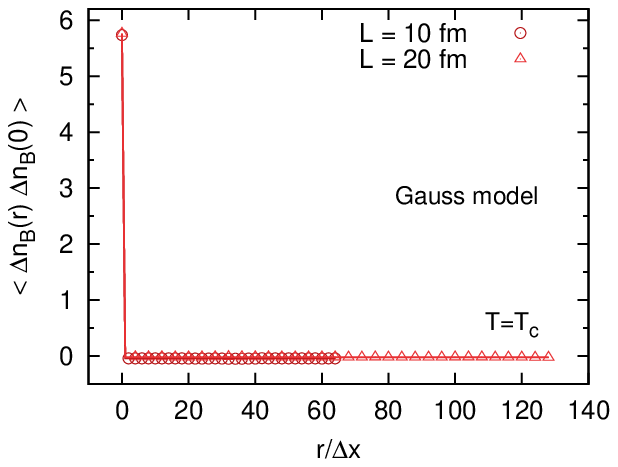}\hfill
 \includegraphics[width=0.48\textwidth]{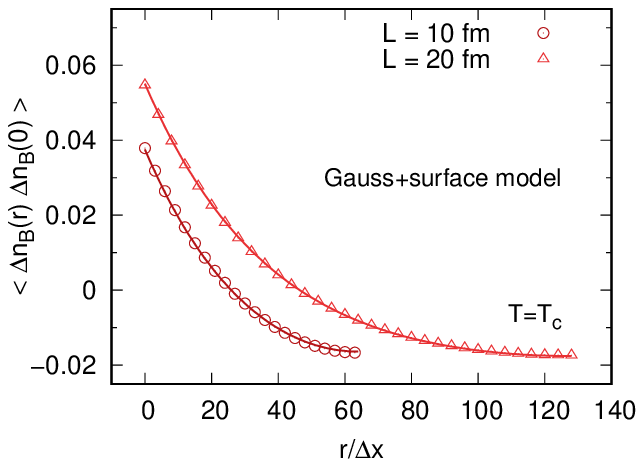}
 \caption{
 Equal-time correlation functions as a function of distance $r$ in units of fixed $\Delta x=L/N_x$ for $T=T_c$ and different $L$ 
 in the Gauss ($K=0$, left panel) and Gauss+surface ($K=1/\xi_0$, right panel) model. The number of grid sites in the numerics 
 (symbols) is $N_x=128$ for $L=10$~fm and $N_x=256$ for $L=20$~fm. The solid curves depict analytic expectations for the employed 
 implicit scheme.}
 \label{fig:correl}
\end{figure}
In Fig.~\ref{fig:correl} we show the equal-time correlation functions $\langle\Delta n_B(r)\,\Delta n_B(0)\rangle$ for the cases $K=0$ 
and $K=1/\xi_0$. For $K=0$, fluctuations are uncorrelated over distances larger than $\Delta x$. Instead, a non-zero surface tension term 
permits the development of real long-range correlations: for $L=20$~fm, we find a numerically realized correlation length of about 
$5.8\,\xi_0$ which goes asymptotically to $\xi(T_c)$ with increasing $L$. In both cases, we observe a non-zero local variance 
$\langle\Delta n_B(0)^2\rangle$ emerging from the purely white noise. While $\langle\Delta n_B(0)^2\rangle_{K=0}$ depends on $1/\Delta x$ in 
line with the thermodynamic expectation $\langle\Delta n_B(r)^2\rangle_{K=0}=(n_c^2/m^2)\,\delta(r)$, for $K\ne 0$ the numerics approaches 
the grandcanonical continuum expectation $\langle\Delta n_B(r)\,\Delta n_B(0)\rangle=(n_c^2/\sqrt{4m^2K})\,e^{-r/\xi}$ as $L\to\infty$. For 
a finite $L$, however, one finds in both cases negative correlations at large $r$ as a consequence of exact charge conservation 
$\int_L\langle\Delta n_B(r)\,\Delta n_B(0)\rangle\,dr=0$ in the finite size system. 
Note that the non-Gaussian fluctuations remain zero for the considered Gaussian models. 

\section{Conclusions}
\label{summary}

The full implementation of intrinsic fluctuations into fluid dynamics presents a challenge for current theoretical models, but is also a 
crucial step toward a realistic description of the QCD dynamics probed in heavy-ion collisions. The accurate treatment of nonlinear 
interactions connected with the necessity of a careful renormalization of cutoff-effects requires a systematic effort. We discussed 
a stochastic diffusion of the net-baryon density which is able to capture the critical phenomena of the Gaussian fluctuations and 
highlights, in particular, the importance of charge conservation in finite-size systems. In future work, we will report on the statics and 
real-time dynamics of non-Gaussian fluctuations~\cite{Nahrgang:VerySoon}. 

\section*{Acknowledgments}
M.B. is funded by the European Union's Horizon 2020 research and innovation programme under the Marie Sk\l{}odowska Curie grant 
agreement No 665778 via the National Science Center, Poland, under grant Polonez UMO-2016/21/P/ST2/04035. This work was financed 
in parts by the U.S. Department of Energy with grants DE-FG02-03ER41260 and DE-FG02-05ER41367. The authors thank Y.~Karpenko for 
providing the vHLLE code and acknowledge fruitful discussions within the Beam Energy Scan Theory (BEST) Topical Collaboration.

\end{document}